\documentclass[onecolumn,showpacs]{revtex4}

\usepackage{graphicx}%
\usepackage{dcolumn}
\usepackage{amsmath}

\makeatletter
\def\btt#1{\texttt{\@backslashchar#1}}%
\DeclareRobustCommand\bblash{\btt{\@backslashchar}}%
\makeatother

\begin{document}


\title{USING  INKSURVEY  WITH  PEN-ENABLED  MOBILE  DEVICES FOR  REAL-TIME  FORMATIVE  ASSESSMENT\\*
II. INDICATIONS  OF  EFFECTIVENESS  IN  DIVERSE  EDUCATIONAL  ENVIRONMENTS\\*
Technology in Practice Strand
}

\author{F.V. Kowalski}
\affiliation{Physics Department, Colorado
School of Mines, Golden CO. 80401 U.S.A.}

\author{Thomas J. Colling, }
\affiliation{Rancocas Valley Regional High School, Mt. Holly NJ, USA}

\author{J. V. Gutierrez Cuba, }
\affiliation{Universidad de las Am\'{e}ricas Puebla, Mexico}

\author{Enrique Palou, }
\affiliation{Universidad de las Am\'{e}ricas Puebla, Mexico}

\author{Gus Greivel, }
\affiliation{Physics Department, Colorado
School of Mines, Golden CO. 80401 U.S.A.}

\author{Todd Ruskell, }
\affiliation{Physics Department, Colorado
School of Mines, Golden CO. 80401 U.S.A.}

\author{Tracy Q. Gardner, }
\affiliation{Physics Department, Colorado
School of Mines, Golden CO. 80401 U.S.A.}

\author{S.E. Kowalski, }
\affiliation{Physics Department, Colorado
School of Mines, Golden CO. 80401 U.S.A.}

\begin{abstract}

{\em InkSurvey} is free, web-based software designed to facilitate the collection of real-time formative assessment.  Using this tool, the instructor can embed formative assessment in the instruction process by posing an open-format question.  Students equipped with pen-enabled mobile devices are then actively engaged in their learning as they use digital ink to draw, sketch, or graph their responses.  When the instructor receives these responses instantaneously, it provides insights into student thinking and what the students do and do not know.  Subsequent instruction can then repair and refine student understanding in a very timely manner.

In a companion paper, we illustrate the wide applicability of this use of technology by reporting a series of seven vignettes featuring instructors of diverse subjects (physics, mathematics, chemical engineering, food science, and biology), with students using diverse pen-enabled mobile devices (tablet PCs, iPads, and Android 4.0 tablets/smartphones), in diverse educational environments (K-12, community college, publicly-funded engineering university, private university, and graduate school), in two countries (United States and Mexico).  In this paper, each instructor shares some data, insights, and/or conclusions from their experiences that indicate the effectiveness of this pedagogical model in diverse educational environments.

\end{abstract}


\maketitle

\section{PROBLEM  STATEMENT  AND  CONTEXT}
\label{sec:problem}

There are broad theoretical foundations for embedding real-time formative assessment in instruction.  However, educators need an efficient, robust method for implementing its collection in the classroom and evidence supporting its effectiveness.

\section{METHOD  EMPLOYED}
\label{sec:method}

{\em InkSurvey} is web-based software designed to facilitate the collection of real-time formative assessment\cite{kowalski2}; it is available for free (\url{http://ticc.mines.edu/}) and is compatible with pen-enabled mobile technology including tablet PCs, iPads, and Android devices (4.0 and higher).   Using this tool, the instructor can embed formative assessment in the instruction process by posing open-format questions.  By avoiding multiple choice questions, this affords more insightful probing of student understanding. Students then use digital ink to draw, sketch, write, or graph their responses, which actively engages them in their learning.  When the instructor receives these responses instantaneously, it provides rich insights into student thinking.  Both the students and the instructor have a more accurate realization of what the students do and do not know.  Subsequent instruction can then repair and refine student understanding in a very timely manner, before misconceptions become deep-rooted, and in a climate where students are receptive to these revisions of their understanding.

\section{RESULTS  AND  EVALUATION:  Seven Vignettes}
\label{sec:results}

There is a growing body of evidence supporting the use of {\em InkSurvey} to collect real-time formative assessment.

\subsection{Evidence of Learning Gains}
\label{sec:learning}

In the companion paper, GG describes in a vignette a lesson using {\em InkSurvey} to reveal and repair a student misconception as they learn about triple integrals over general regions of three-dimensional space. When 87 students were examined in a summative assessment two weeks after the lesson, only one student still displayed this mistake on any of the three exam questions that addressed the concept.  By way of comparison, in GG's other large lecture section of Calculus III, in which {\em InkSurvey} was not being used, this particular misconception was given similar consideration in lecture and $12$ out of $103$ students made this mistake on the same set of exam questions.

Similarly, TC's high school students, who receive regular and repeated individual and whole-group feedback as part of instruction using {\em InkSurvey}, have gained traction on mastery of fundamental concepts in Algebra I, as measured by standards-based assessments and in comparison to students who were not exposed to the {\em InkSurvey} intervention \cite{maniglia}. Interestingly, even though his classes are mixed gender, female students in particular demonstrated much greater learning gains than the control group. Student attitudes toward learning with {\em InkSurvey} are discussed later in this paper, but it is noteworthy here that when TC's students were surveyed, the females had a much stronger agreement (70\%) vs. males (48\%) with the statement: ``{\em InkSurvey} helped me to know what I understood of the topic and where I needed further help."  Since these observed differences in both attitude and learning gains in females vs. males could have significant implications, particularly in STEM education, further investigation is warranted.

In Food Chemistry courses at Universidad de las Am\'{e}ricas Puebla, redesigned to incorporate {\em InkSurvey} and other research-based pedagogy, EP, JC, and collaborators investigated changes as frequent formative assessment helped make students' thinking visible to themselves, their peers, and their instructor. They documented increased student participation in class discussions and problem-solving activities \cite{cuba1}, while instructors utilized the information gained through real-time formative assessment to tailor instruction to meet student needs. Through qualitative and quantitative analysis of the information obtained from six semesters, they report the impact of creating classroom tasks and conditions under which student thinking can be revealed \cite{cuba1,cuba2,palou}.

In both undergraduate and graduate courses, formative assessment exercises performed with Tablet PCs and {\em InkSurvey} had a positive impact on performance on a series of summative quizzes. These summative assessments were compared with performance of students in the same courses four years before implementing these revisions, and showed mean improvements of $0.6$ (undergraduate course) and $0.5$ points (graduate course) out of $10$ possible \cite{cuba2}. The frequent formative assessments using {\em InkSurvey} at UDLAP generated possibilities for self-assessment, allowing students to anonymously analyze their own and classmates' thinking. Other important impacts this team observed include:  ability of instructor to identify the most common difficulties and provide immediate feedback; student reflection on their own processes as learners; and improvements in both classroom instruction and student academic success when the instructor has these insights into student thinking \cite{cuba1,cuba2,palou}.

At Colorado School of Mines, {\em InkSurvey} has also been used to strengthen problem-solving skills in an upper-level engineering physics course \cite{kowalski3}.  As $11$ new problems were introduced throughout the course, students, working individually or in groups, provided real-time formative assessment of their problem-solving skills by submitting responses to a standard series of three questions (designed as nearly universally applicable in problem solving). These responses were later analyzed to determine the students' ability to apply the problem-solving strategy; scores were adjusted to reflect the difficulty of the problems.  Results show a steady improvement over the semester in problem-solving skills, indicating {\em InkSurvey} can potentially nurture higher level thinking skills.

When {\em InkSurvey}'s real-time formative assessment is coupled with interactive computer simulations (sims), strong learning gains are achieved. In two Chemical Engineering courses, TG targeted $6$ concepts for which students in the past have had difficulties visualizing the connections between the calculations and the physical processes \cite{gardner}. In this study, student understanding was compared at 3 points in time during the learning process: before a sim was introduced, after the students had played with the sim on their own, and after the instructor used {\em InkSurvey} to probe student understanding and guide their further explorations of the sim.  After playing on their own with the sims, and knowing what question would be asked at the end, students' average level of competence across the six topics in this study increased from  $\sim 1.8$ to $\sim 2.3$ (on a $4$ point scale), indicating students still did not adequately understand the concepts after exploring the simulations in an unguided manner.  Without the instructor ever telling students the answers to the questions, but instead posing scaffolded questions based on the students' immediate issues and misconceptions and allowing students to further explore the sims as they answered these questions, students' understanding of these concepts increased to an average of $\sim 3.1$ on the $4$ point scale on these same topics.

Our final evidence in this section addresses the effectiveness of using {\em InkSurvey} for real-time formative assessment with students of different learning styles.  FK and SK hypothesized that perhaps the graphical nature of {\em InkSurvey} input makes it more appealing or effective for learners with strongly visual or kinesthetic learning styles.  However, their data reflects strong learning gains achieved when {\em InkSurvey} is used for students of all learning styles \cite{kowalski4}. This surprising result is particularly encouraging when considering implementation in classes with typical diversity among students.

\subsection{Student Insights}
\label{sec:insights}

Student response to the use of {\em InkSurvey} has in general been very favorable.  In the Fall $2012$ semester, FK used {\em InkSurvey} in an Advanced Laboratory class for junior-level undergraduate physics students.  In an anonymous survey of $63$ students, $54$ students (86\%) agreed that their responses with {\em InkSurvey} gave the instructor a better understanding of what they do and do not understand.  Some of their comments include:

\begin{itemize}
\item ``It gives instant feedback to him so he can see right away if what he just talked about made any sense. If it did, we can move on, if not, he re-explains it in a different way or elaborates on it more which is good."
\item ``I do think {\em InkSurvey} has let the professor understand what we know better because it is an
efficient way to test us. This quick response lets him see what topics the class is struggling with and what topics we get based on the answers he gets back. If there are a wide variety of responses, we are probably struggling with something. When this happens, we go over the topic slowly so we can understand it better. If there is a common wrong answer, we can go over the mistake made to get there and prove why it is wrong. If most the class is able to get something correct, we spend very little time discussing what the right answer is because we already get it."
\item ``Yes, I feel that they give the instructor a better understanding of what I do and do not understand because he is able to look directly at my work and follow my thinking. Every time that I have answered incorrectly or been missing a portion of the question, he has gone over it after he has looked over the {\em InkSurvey}."
\item	``I feel that my responses using Ink Survey have given the instructor a better understanding of
what I know or don't know. In most classes student understanding is conveyed through homework and quizzes. This is great but by the time quizzes and homeworks are graded the class has moved onto new topics and it is very inconvenient for the class to slow its momentum and revisit old material. Ink Survey is a great resource for a teacher because it gives instantaneous feedback to the teacher when a topic is being introduced. Because of this the professor can change his or her teaching style based on the class understanding."
\end{itemize}

In the same survey, a question was asked to probe the effectiveness of this pedagogical model in increasing student metacognition. Forty-eight students (76\%) found that creating their responses in {\em InkSurvey} actually helped them better understand what was being covered in class, or helped them realize what they did and did not understand.
\begin{itemize}
\item ``Having to physically write down my response to all the questions is very helpful and shows me where I am going wrong or what I do not understand so that I may remedy it."
\item	``Answering questions about what one has just learned often increases, at the very least, your memory of an event. Further, if you didn't understand something, having a fast reiteration of the correct answer is conducive to comprehension. In this regard, {\em InkSurvey} works well because it is faster than a paper quiz and can have everyone answer independently, unlike an oral exchange."
\item	``I believe that Ink Survey is extremely helpful in the understanding of class topics. This is because it keeps students in the classroom involved and participating in the classroom. Using traditional teaching methods students are not forced to think about what is being taught and often get in a routine of copying notes from the board to paper and then thinking about the topic later when doing homework. {\em InkSurvey} forces you to stay involved in the classroom because it allows the professor to ask thought provoking questions, making you think about what is being taught at that moment--not later after class when doing homework. Because of this more questions are brought up in class, facilitating the learning environment that {\em InkSurvey} brings."
\item	``It sometimes also makes me realize I know something better than I thought I did or I already knew how to do it and that it's not as scary as it may seem when it's presented.
\item	``I think having to draw out and write the solutions shows me my shortcomings with the material very clearly. You cannot fake knowing an answer when you have to draw things out."
\end{itemize}

Focus group conversations revealed that sometimes students resent the demand {\em InkSurvey} places on them to participate in class, rather than passively sitting there. However, in an introductory physics class, TR's students indicate they feel much more engaged with open-format questions than with multiple-choice clicker questions.  Because students can't just select an answer, they feel more pressure when writing something because, as one student explained, they ``don't want to write something stupid."  While they understand that it's acceptable to submit an incorrect response, their self-imposed pressure encourages students' additional engagement with the question. Another physics student at Colorado School of Mines feels that needing to quickly think through a problem and construct a viable solution to submit on {\em InkSurvey} is providing him/her with excellent preparation for future job interview tasks and on-the-job challenges.

At Universidad de las Am\'{e}ricas Puebla, graduate students using {\em InkSurvey} felt that this: increased their motivation to participate in class as well as their scores in graded work-products; made the classroom more active and kept them constantly thinking, thus increasing their learning with understanding; enabled the teacher to provide a great deal of real-time feedback to students that made their thinking visible; and gave them chances to revise their understanding [4].

\subsection{Instructor Insights}
\label{sec:instructor}

Instructors agree that there is an art to effectively using {\em InkSurvey} for real-time formative assessment, both in construction of appropriate questions and in agilely responding to the student submissions \cite{kowalski5}.  Two often opposing needs must be balanced. On one hand, there is a need to take full advantage of the open-format questions and allow students enough room to construct correct understanding and demonstrate higher levels of thinking.  On the other hand, if questions are too complex or lengthy, the instructor risks losing the interest of students whose understanding is not yet mature enough to tackle the question. With practice, instructors learn to break questions into appropriate parts.

There is also some skill, which again improves with practice, in knowing how to phrase questions to deeply probe student understanding and yield responses that give clear insights into student thinking.  A picture is worth a thousand words, and the power of responses constructed with digital ink is their ability to display a wealth of conceptual understanding in a graph, a sketch, a Venn diagram, the mathematical solution to a problem, or the outline of a proof.  These graphical responses can be quickly scanned by the instructor as they are received and can effectively serve as the foundation of subsequent class discussion.  In contrast, this is probably not an effective format for soliciting from the students routine five-paragraph essays defending a particular point of view, or an epic poem in stanzas of iambic pentameter, since real-time formative assessment is neither as useful nor as practical in these cases.

Some instructors implementing the collection of real-time formative assessment use {\em InkSurvey} in a controlled online environment.  However, many others are acutely aware of the temptations to students when they have ready access to the internet during class.  One way to address this classroom challenge is to anticipate it and prepare to further engage the faster workers as they wait for their peers to submit their responses.  {\em InkSurvey} allows the instructor to independently launch access to any question at any time. FK sometimes employs this to facilitate differentiated instruction and keep every student on task. He starts the entire class working on a question (or series of questions) on a fundamental concept; as he begins receiving correct responses from students, he then launches another question for them to address.  This is often an enrichment question, encouraging deeper student understanding.  By the time the majority of students have responded to the original question, a subset of them will have also considered the enrichment question, and some of that group will have submitted responses.  Even though not all students will have individually experienced the enrichment question, the responses received can serve as the foundation for class discussion of this after the instructor responds to the real-time formative assessment received from the original question related to the fundamental concept \cite{kowalski5}.

\section{CONCLUSION}
\label{sec:concl}

A companion paper \cite{kowalski1} describes the use of {\em InkSurvey} for real-time formative assessment to enhance learning; a growing body of data, summarized here, supports the effectiveness of this pedagogical tool. This includes evidence of learning gains, positive student attitude feedback, and instructor insights from a variety of educational environments and using a variety of pen-enabled mobile devices (tablet PCs, iPads, and Android $4.0$ tablets).

\begin{acknowledgments}
Various facets described here have been supported by:  HP Catalyst Program (FK, SK, TC, JC, TG, and EP), NSF grants \# 1037519 and \# 1044255 (FK, SK, and TG), the Trefny Endowment (GG, TR), and the National Council for Science and Technology of Mexico (CONACyT) (JC).  We appreciate this generous support.
\end{acknowledgments}

\end{document}